\documentclass[a4paper]{spie}  
\usepackage[]{graphicx}

\title{An apodizing phase plate coronagraph for VLT/NACO}

\author{Matthew A. Kenworthy\supit{a,b}, Sascha P. Quanz\supit{c},
Michael R.  Meyer\supit{c}, Markus E. Kasper\supit{d}, Rainer Lenzen\supit{e},
Johanan L. Codona\supit{b}, Julien H.V. Girard\supit{f},  and
Philip M. Hinz\supit{b}
\skiplinehalf
\supit{a}Leiden Observatory, Leiden University, P.O. Box 9513, 2300 RA Leiden, The Netherlands; \\
\supit{b}Steward Observatory, 933 N. Cherry Avenue, Tucson, AZ 85721, USA; \\
\supit{c}Institut f\"ur Astronomie, ETH Z\"urich, Wolfgang-Pauli-Strasse
27, 8093 Z\"urich, Switzerland; \\
\supit{d}European Southern Observatory, Karl-Schwarzschild-Str. 2, D-85748 Garching, Germany; \\
\supit{e}Max-Planck-Institut f\"ur Astronomie, K\"onigstuhl 17, 69117 Heidelberg, Germany; \\
\supit{f}European Southern Observatory, Casilla 19001, Santiago, Chile\\
}

\authorinfo{Send correspondence to M.A.K. E-mail: kenworthy@strw.leidenuniv.nl, Telephone: +31 (0)71 527 8455}
\begin{document}
  \maketitle 

\begin{abstract}

We describe a coronagraphic optic for use with CONICA at the VLT that
provides suppression of diffraction from 1.8 to 7 $\lambda/D$ at 4.05
microns, an optimal wavelength for direct imaging of cool extrasolar
planets. The optic is designed to provide 10 magnitudes of contrast at
0.2 arcseconds, over a ``D'' shaped region in the image plane, without the
need for any focal plane occulting mask.

\end{abstract}

\keywords{coronagraph, high contrast, adaptive optics, thermal infrared}

\section{INTRODUCTION}
\label{sec:intro}  

Many telescopes are being used in the quest for direct imaging of
extrasolar gas giant planets around nearby stars. Detecting the signal
of an extrasolar planet directly is challenging due to the large contrast
ratio expected between the planet and star. To date, most surveys
for extrasolar planets have obtained null results constraining the
frequency of young gas giants at large orbital radii ($>$ 30 AU).
However, recent notable successes
\cite{Marois2008,Kalas2008,Lagrange2009,Lagrange2009B,Lafreniere2010,Lagrange2010}
have provided evidence for rare, yet remarkable systems that challenge
our ideas of planet formation. Based on our understanding of circumstellar
disks that surround young sun-like stars, as well as hints from radial
velocity surveys, we expect most planet formation to occur between
$3-30$ AU.

To this end, coronagraphs are employed to minimise the scattered and
diffracted flux from the host star that can obscure the signal from
the exoplanet. Other techniques, such as SDI, have lower throughput
since narrow band filters are utilised, and depend on the veracity
of atmospheric models, requiring a methane absorption feature $(T_{eff}>1200K)$
to distinguish the planet from the host star.

Imaging in the thermal IR from $3-5\mu m$ enables searches for colder
planets. Consequently, surveys can concentrate on older stars that
tend to be closer to the Sun, permitting searches at smaller physical
separations comparable to the dimensions of our Solar System when
employed to study the nearest stars. Using an APP
coronagraph \cite{Kenworthy2007}, broadband observations at longer
wavelengths (where the planet/star contrast ratio is more favourable)
such as L and M band, in combination with the AO system at the VLT,
will enable exoplanet searches at small ($<0.5$ arcseconds) angular
separations corresponding to $3-30$ AU around nearby candidate stars
($6-60$ pc).

As we are just on the verge of many new
exoplanet detections, we believe this factor of three will yield an
unanticipated number of new discoveries. In addition, this capability
can be used to improve characterization of objects discovered at other
wavelengths, improving constraints on the physical characteristics
of planets from broader wavelength coverage. Finally, we speculate
that the thermal IR will emerge as the wavelength of choice to study
the formation of terrestrial planets through detection of hot protoplanet
collision afterglows such as predicted from models of the formation
of the Earth-Moon system
\cite{Stern1994,Mamajek2007,Zahnle2007,Miller-Ricci2009}. 

The APP coronagraph has broad applicability, useful for imaging and characterization
of any astronomical object with faint extended structure surrounding
bright targets, including, but not limited to, extrasolar planets. 

\section{PRINCIPLE OF THE APODIZING PHASE PLATE}

\subsection{Overview}

\begin{figure}
\begin{center}
\includegraphics[angle=270]{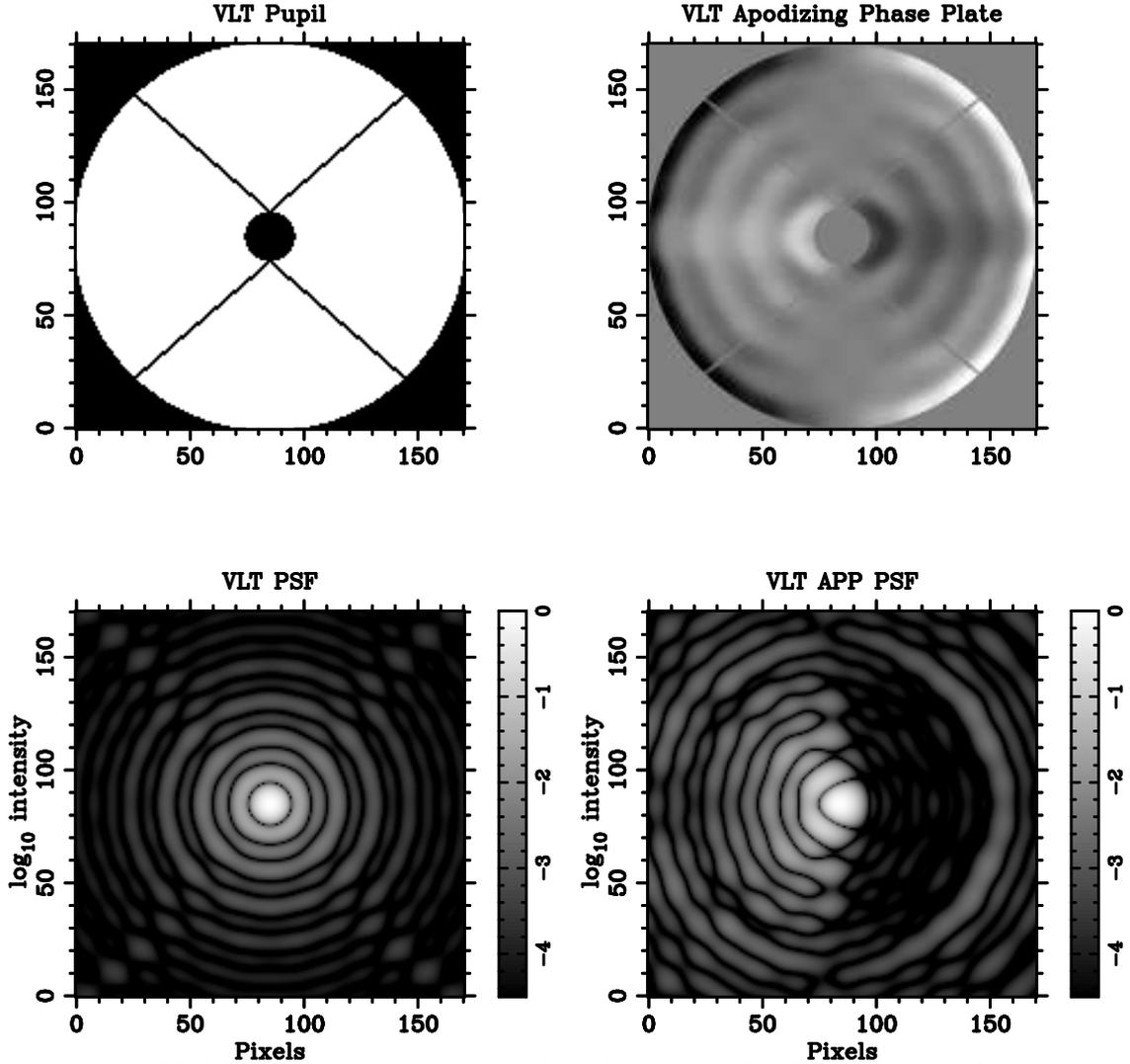}
\caption{\label{fig:VLT-APP}The VLT APP Design. The upper two panels
show the VLT pupil and the phase modification that the APP optic introduces
to the VLT pupil. The resultant PSFs are shown in the lower two panels,
with a logarithmic intensity scale normalised to the peak flux in
each case. The encircled energy flux for the APP is 59\% of the unobstructed
VLT PSF flux at $3-5\mu m$.}
\end{center}

\end{figure}

\begin{figure}
\begin{center}
\includegraphics[angle=270]{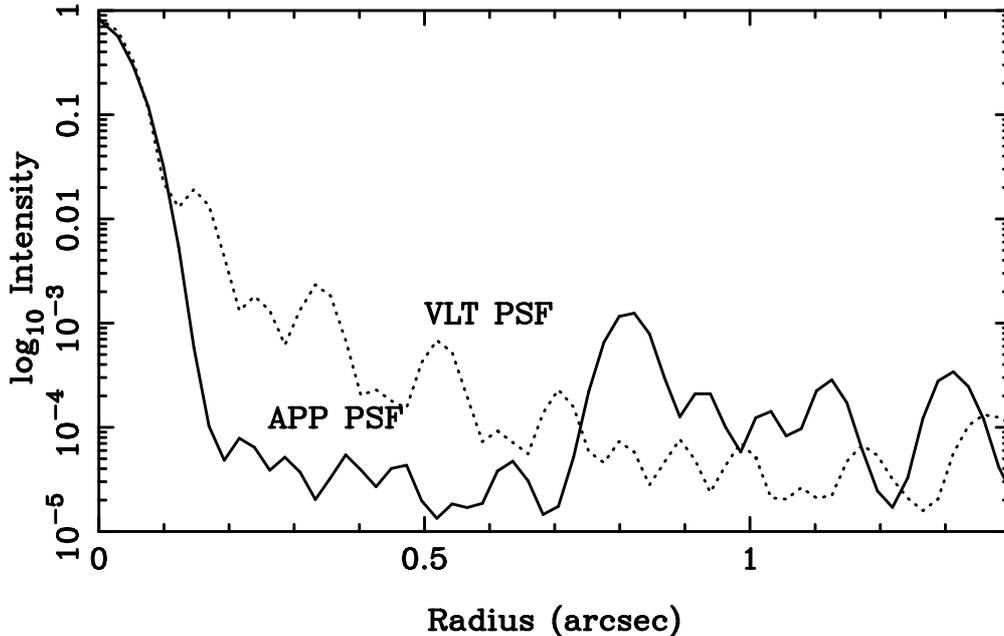}

\caption{\label{fig:Azimuthally-averaged-flux}Azimuthally averaged flux of
the VLT and APP point spread functions from Figure \ref{fig:VLT-APP}.
Average over a wedge $150^{o}$ centered over the dark D of the APP.}
\end{center}

\end{figure}

Like many other telescopes and imaging cameras, CONICA has a pupil
filter wheel that is coincident with the location of the reimaged
telescope pupil. At this point in the optical train, the pupil image
is at a location where the wavefront from a distant astronomical object
is nominally flat and planar. The resultant image at the science camera
has the typical point spread function (PSF) of a large, adaptive optic
corrected telescope. By modifying the wavefront phase at this pupil
plane, this Airy ring pattern can be rearranged and diffraction rings
suppressed over a $180^{o}$ field of view ``D'' shaped region
\cite{Kenworthy2007,Codona2006},
see Figure \ref{fig:VLT-APP}. The reduction in Airy ring flux is shown
in Figure \ref{fig:Azimuthally-averaged-flux} where the effect of the
APP shows suppression by a factor of 10 to 50.

\subsection{Resultant Point Spread Function (PSF) using the APP}

All coronagraphs suffer from a loss of sensivitity, due to the scattering
of science light through the optics. Focal plane masks block out the
central star and an extended region surrounding it, resulting in a
lack of close-in sensitivity and a loss of angular resolution when
a Lyot stop is used in the pupil plane. Furthermore, there is a cost
in the overhead used in aligning the coronagraph with the host star.

The APP is different because there is no amplitude apodization - all
of the flux of the pupil is transmitted through the APP, only the
phase is modified. Since we are using light from the core of the PSF
to suppress diffraction over one half of the PSF, there is a reduction
in the core intensity of all objects in the field of view of the camera,
including the planet and host star. The degree of core flux lost is
a trade-off between level of suppression, the smallest inner working
angle that the suppression can go to, and the core flux \cite{Codona2007}.
We choose an (APP/Direct Imaging) core flux ratio of 59\% (see Figure
\ref{fig:VLT-APP}). The integration required to get complete coverage
around a host star and reach the same signal to noise as direct imaging
is 4 times longer, but this is more than compensated for by the \textbf{additional
sensitivity at small IWAs which are otherwise inaccessible without
a coronagraph.}

\subsection{Realization of the APP Coronagraph}

Variations in the thickness of a Zinc Selenide substrate introduces
optical path differences that correspond to variations in phase across
the telescope pupil. The variation in phase, $\Delta\phi$, is related
to variations in the thickness of the substrate $\Delta x$ by the
refractive index $n$ of the transmitting material: $\Delta\phi=(2\pi/\lambda).\Delta x.(n-1)$,
which is an inherently chromatic process. However, the performance
of the APP degrades only slowly with increasing bandwidth and the
plate is used at 20\% bandwidth at the MMT from $3-5\mu m$.

\begin{figure}
\begin{center}
\includegraphics[scale=0.5,angle=0]{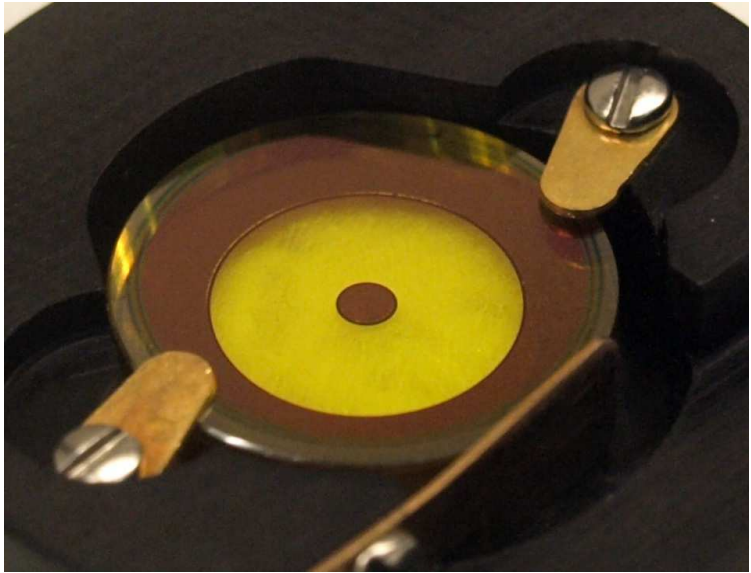}

\caption{\label{fig:app_photo}
Photograph of VLT APP Optic mounted in holder. The light yellow area is
anti-reflection coated Zinc Selenide, the darker areas are gold coated
to prevent stray light from a misaligned pupil passing through the
plate.}
\end{center}

\end{figure}

The VLT APP is manufactured in a multi-stage process. First, a blank of
Zinc Selenide has a pedestal with the same dimensions to the telescope
pupil machined into its surface. A gold coating is deposited over the
side of the optic with the pedestal, and then the APP phase pattern is
machined into the pedestal, removing the gold mask in the process. The
optic is then anti-reflection coated for a central wavelength of
$4.05\mu m$. The final optic is shown in Figure \ref{fig:app_photo}.

\section{FIRST LIGHT PERFORMANCE AT THE VLT}

We now summarize the results of the commissioning run for the
APP coronagraph that was recently installed in
the CONICA camera mounted at VLT UT4. The commissioning tests were
carried out on April 2 and 3, 2010. The objectives were to characterize
the performance of the APP so that observers can accurately plan
and carry out their observations. All tests were made with the narrow
band filter NB4.05 or the L$'$ filter ($\lambda_{\rm cen}=3.8\,{\rm \mu
m}$) using the L27 camera with a pixel scale of $0.027''$/pixel.
The new intermediate band filter IB4.05 could not be tested with the APP
due to difficulties obtaining a proper focus.

\section{Alignment of the APP with the pupil}

\begin{figure}[t]
\begin{center}
\includegraphics[scale=0.6,angle=0]{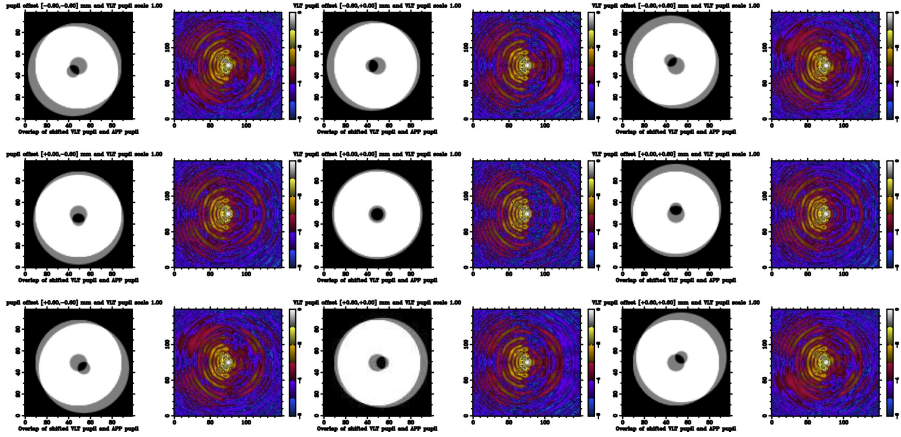}

\caption{Simulations showing the effect of APP-pupil misalignment on the
predicted PSF. The center image pair shows a perfectly aligned APP
(white) with the VLT pupil (grey) and an almost perfectly clean
hemisphere on the righthand side of the APP PSF where the diffraction
pattern is effectively suppressed. The other image pairs show the result
from offsets of 60\,$\mu$m between the APP and the VLT pupil in positive
and negative x and y direction and combinations thereof. The
effectiveness  of the diffraction ring suppression is worse in all
cases. }

\end{center}
\label{misalignment}
\end{figure}

In order to take full advantage of the APP capabilities it is necessary
that the APP is well aligned with the pupil. As shown in
Fig.~\ref{misalignment} a misalignment between the APP and the pupil
leads to imperfect suppression of the diffraction rings in the normally
``clean'' part of the PSF between $\sim$0.2--0.7$''$. As the APP sits in
the Lyot wheel there is only one remaining degree of freedom (the
rotation of the wheel) that can be changed in order to achieve best
possible alignment. Changing the position of the APP in the wheel in
radial direction with respect to the Lyot wheel center is not possible. 

\begin{figure}[t]
\begin{center}
\includegraphics[scale=0.5,angle=0]{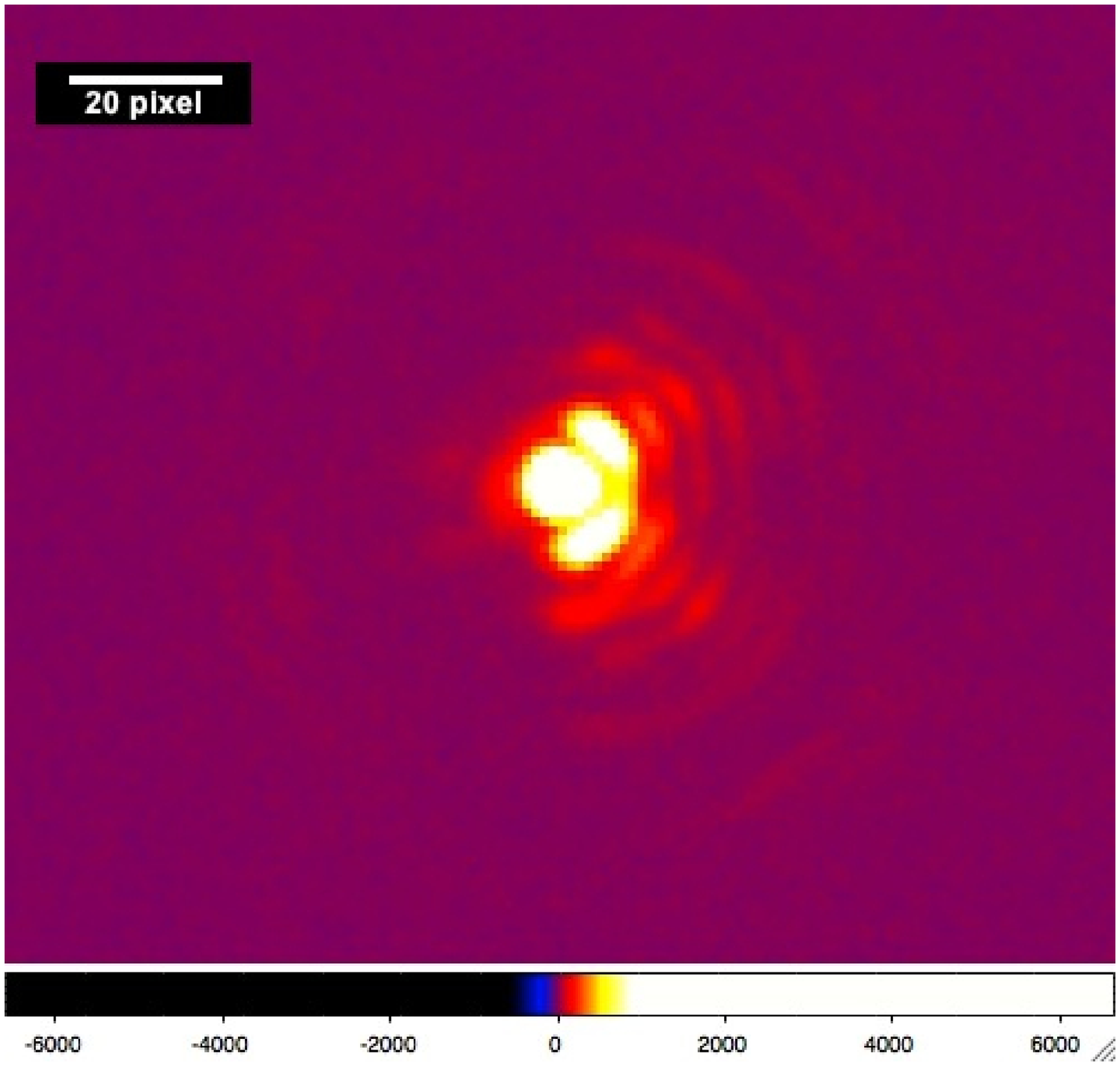}

\caption{NACO/APP PSF after the alignment test. The ``clean'' part of the
PSF is seen in the left hemisphere where the diffraction pattern is
suppressed compared to the non-corrected hemisphere. The image is a
single 30 sec exposure (DIT = 1 s, NDIT = 30) of HIP39156 in the NB4.05
filter. Dark current and sky background were eliminated by subtracting
the consecutive image of the dither pattern from this exposure.}

\label{PSF}
\end{center}
\end{figure}

The first test during the commissioning run was to ensure that the
encoder position of the Lyot wheel when the APP is inserted in the light
path corresponds to the best possible alignment between APP and pupil.
For this purpose the bright standard star HIP39156 (K0 III, L$'$ = 5.1
mag) was observed in the NB4.05 filter (DIT = 1 s, NDIT = 30, NINT =
5)\footnote{DIT = "Detector Integration Time", NDIT = Number of coadds
with same DIT in a single file; NINT = the total number of dither
positions for which exposures of DIT$\times$NDIT s were obtained.} with
different encoder positions. The resulting images were compared to
simulations in order to determine any possible offsets between the APP
and the pupil. In total 10 different encoder positions were analyzed and
the best results were obtained for an encoder position of 48595. This
value was taken as the default encoder position for the rest of the
commissioning run. Figure \ref{PSF} shows the corresponding PSF.

\section{Throughput measurements}

The APP takes energy from the core of the PSF and redistributes it so
that the diffraction pattern in one hemisphere is suppressed while it is
enhanced in the other hemisphere. As a result, the flux in the PSF core
is lower compared to non-APP observations. This flux reduction is
quantified below in order to help the planning of APP observations.

\subsection{Throughput in NB4.05 filter}
\subsubsection{Observational setup}

For this test the bright standard star HIP52804 (K0 III, L$'$ = 4.3 mag)
was used.  First, a set of exposures was taken without the APP, then an
equivalent set with the APP. The time lag between the exposures without
and with the APP was roughly 10 minutes. The exposures had DIT = 0.35 s,
NDIT = 30, and NINT = 5 and were repeated three times yielding in total
15 individual images each having $\sim$10.5 s integration time. The
airmass was $<$1.1 and the DIMM seeing was $\sim$0.9$''$ and
$\sim$0.85$''$ without and with the APP, respectively.

\subsubsection{Basic data reduction steps}

Bad pixels deviating by 3-$\sigma$ from the mean of a surrounding 5
pixel box were replaced by the mean of the neighboring pixels. Dark
current and sky background were eliminated by subtracting two
consecutive images taken at different dither positions from each other.
The remaining background level was always close to zero. 

\subsubsection{Throughput analysis}

To determine the throughput of the APP, the count rate in the inner
pixels of the PSF in each individual image was computed using the IDL
routine {\tt ATV.pro}. Two different aperture sizes were used (2 pixel
radius and 5 pixel radius). Table 1 summarizes the results. The count
rate is the mean of the 15 individual images and the error is the
corresponding standard deviation of the mean value. The throughout is
simply the ratio of the count rates with and without the APP. It shows
that for the NB4.05 filter the throughput in the PSF core is  59\% and
63\% for an aperture with a radius of 2 and 5 pixels, respectivly. This
is in very good agreement with theoretical predictions that estimated a
value of $\sim$56\% in the innermost regions of the PSF core.

\begin{table}[h]
\caption{Throughput measurements for the VLT APP as measured in April
2010.}
\label{tab:fonts}
\begin{center}       
\begin{tabular}{|l|l|l|} 
\hline
\rule[-1ex]{0pt}{3.5ex}   & 2 pixel & 5 pixel \\
\hline
\rule[-1ex]{0pt}{3.5ex}  NB 4.05 filter & $0.59\pm0.04$ & $0.63\pm0.03$ \\
\hline
\rule[-1ex]{0pt}{3.5ex}  L$'$ filter     & $0.55\pm0.04$ & $0.61\pm0.03$ \\
\hline
\end{tabular}
\end{center}
\end{table} 

\subsection{Throughput in L$'$ filter}

For the L$'$ filter  ($\lambda_{\rm cen}=3.8\,{\rm \mu m}$) the bright
standard star HIP61460 (F2 V, L$'$ = 6.2 mag) was used.  Again, first a
set of exposures was taken without the APP, then an equivalent set with
the APP. The time lag between the exposures without and with the APP was
roughly 10 minutes. The exposures had DIT = 0.175 s, NDIT = 30, and NINT
= 5 and were repeated three times yielding in total 15 images each
having $\sim$5.25 s integration time. The airmass was $<$1.1 and the
DIMM seeing was $\sim$0.6-0.8$''$ for all exposures. The basic data
reduction and throughput analysis was the same as described in the
previous sections.

Table 1 summarizes the results for the L$'$ filter. The throughput in
the PSF core is roughly 55\%, again in agreement with the models.

\section{Contrast curve measurements}

\subsection{Contrast curve in the L$'$ filter}

\subsubsection{Observational setup}

HIP61460 (F2 V, L$'$ = 6.2 mag) was used for the test in the L$'$ filter
($\lambda_{\rm cen}=3.8\,{\rm \mu m}$). The observations were done in
pupil tracking mode with DIT = 0.175 s, NDIT = 30, and NINT = 5 yielding
5 images at different dither positions. To obtain in total roughly 30
minutes on source integration time this observing template was repeated
70 times yielding in total 350 individual image/files. All images were
in the linear detector regime ($\sim\frac{1}{3}$ Full Well). The DIMM
seeing varied typically between $\sim$0.7--0.8$''$ and the airmass was
$<$1.1.

\subsubsection{Basic data reduction steps}

Bad pixels deviating by 3-$\sigma$ from the mean of a surrounding 5
pixel box were replaced by the mean of the neighboring pixels. Dark
current and sky background were eliminated by subtracting two
consecutive images from each other. The remaining background level was
always close to zero. Afterwards, all images were aligned with an
accuracy of 0.1 pixel using cross-correlation. 

\subsubsection{Computation of contrast curve and results}

All images were stacked together and a ``MEAN image'' and a ``RMS
image''
were created by computing the mean and the r.m.s. of each pixel in the
stack.  Then, the mean flux per pixel in the core of the PSF was
computed in an aperture with a 5 pixel radius.

\begin{equation}
\bar{f}_{core}=\frac{\sum_{i=1}^{n}f_i}{n}
\end{equation}

with $f_i$ being the flux in the individual pixels and $n$ being the
number of pixel considered ($5^2\cdot\pi\approx 78$ in our case). The
``RMS image'' was divided by the square root of the number of individual
images in the stack, i.e., 350. The resulting ``One sigma image'' ($
\rm{Im_{1\sigma}}$) contains the one sigma noise for a single pixel
planet. To compute the contrast as a function of radius, we have to take
into account that the flux of a potential companion will be distributed
across several pixels and we have to use the same 5 pixel aperture that
we used for the PSF core, i.e., $n$ pixel. This will increase the signal
to noise by a factor $\sqrt{n}$. Thus, the ``5--$\sigma$ Contrast
image''
in units of magnitudes can be computed using

\begin{equation}
{\rm Im_{5\sigma}}=-2.5\cdot{\rm log}\Bigg(\frac{5\cdot \rm{Im_{1\sigma}}}{\sqrt{n}}\frac{1}{\bar{f}_{core}}\Bigg)\quad .
\end{equation}

\begin{figure}[t]
\begin{center}
\includegraphics[scale=0.63,angle=270]{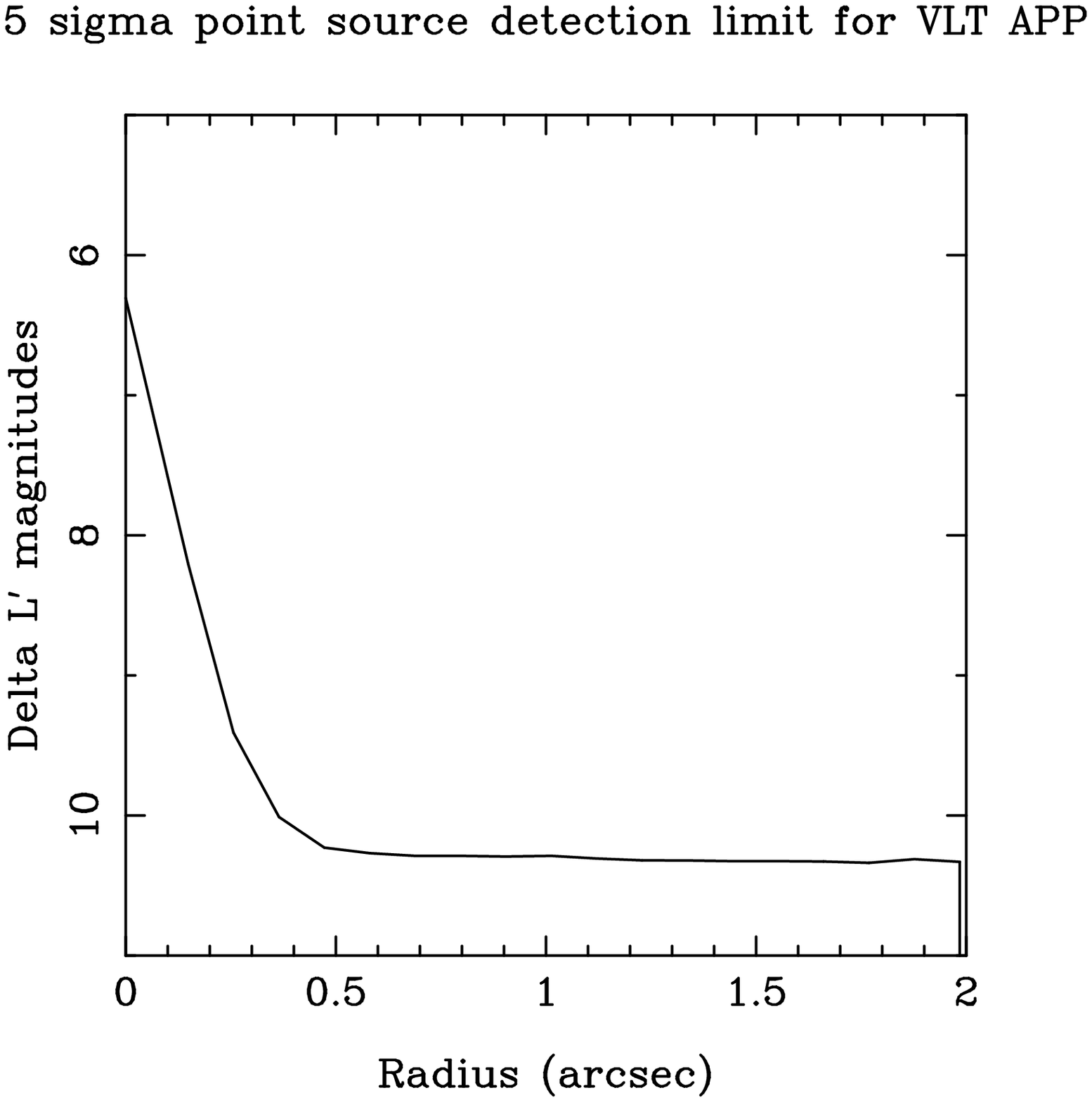}

\caption{NACO/APP contrast curve for HIP61460 in the L$'$ filter
($\lambda_{\rm cen}=3.8\,{\rm \mu m}$) and the observational setup
described in Section 6.1.1. At $\sim$0.5$''$ the sky background limit is
reached. }

\label{contrast_image}
\end{center}
\end{figure}

In this image each pixel contains an estimate of the delta magnitude of
a 5--$\sigma$ point source detection with respect to the central core
PSF. Fig.~\ref{contrast_image} shows the corresponding contrast curve
for HIP61460. The contrast curve was derived  in the ``clean'' hemisphere
of the PSF in a wedge with a 160$^{\circ}$ opening angle. The values
plotted in Fig.~\ref{contrast_image} correspond to the mean value of
4-pixel wide semi-annuli in this wedge for a given radial distance.

\subsection{Discussion of the contrast curve}

The contrast curve shown in Figures~\ref{contrast_image} show that the
APP is working as expected and approximately as predicted from
simulations. The steep drop of the curve at very small inner working
angles demonstrates the specific strength of the APP, namely enhancing
the achievable contrast close to the central star compared to non-APP
images. 

Although we took contrast curve data for the APP in the NB4.05 and
IB4.05 filters, we did not get enough data to explore the contrast
limits of the phase plate. We reach the background limit for the L$'$
filter at 0.5$''$.  The APP is specifically designed to achieve the best
performance at $4.05\mu m$ with the NB4.05 filter and the new IB4.05
filter. However, our tests have demonstrated that the APP performs very
well with the L$'$ filter as well, and since the bandwith for the NB and
IB filters are narrower than that for the L$'$ filter, we expect that
the performance will be at least comparable to or better than the L$'$
filter contrast curve.

\section{CONCLUSIONS}

In general, the APP is expected to work best (in terms of contrast) around 4
$\mu$m in combination with the NB4.05 and with the IB4.05 filter.
Our results with the L' filter confirm that the APP is working well, and
expected to obtain higher contrasts at narrower bandwidths.

Observing with the APP should not increase the overhead that has
to be considered for the planning and execution of observations apart
from the overhead resulting from the reduced coronagraphic throughput of
$\sim$59\%, which should be taken into account in the exposure time
calculations. The APP sits in the pupil plane and can be used in any of
the common imaging modes (e.g., pupil tracking) once it is rotated in
the light path. Obviously one has to take into account, however, that
only one hemisphere around a target shows an improved contrast behavior
so that a second data set (with the FoV rotated by 180 deg) is required
in order to cover both hemispheres around a target.

\bibliographystyle{spiebib}   
\bibliography{mn-jour,newkenworthy}

\end{document}